# Single-shot measurement of space varying polarization state of light through interferometric quantification of geometric phase

Athira B S[1,*], Mandira Pal[2,*], Sounak Mukherjee[2], Jatadhari Mishra[2], Dibyendu Nandy[1,2] and Nirmalya Ghosh[1,2]

[1]*Center of Excellence in Space Sciences India, Indian Institute of Science Education and Research Kolkata, Mohanpur 741246, West Bengal, India*

[2]*Department of Physical Sciences, Indian Institute of Science Education and Research Kolkata, Mohanpur 741246, West Bengal, India*

[*]*Corresponding authors:* abs16rs013@iiserkol.ac.in, mandirapal89@gmail.com

**Abstract**

Light beam carrying spatially varying state of polarization generates space varying Pancharatnam-Berry geometric phase while propagating through homogeneous anisotropic medium. We show that determination of such space varying geometric phase provides a unique way to quantify the space varying polarization state of light using a single-shot interferometric measurement. We demonstrate this concept in a Mach-Zehnder interferometric arrangement using a linearly polarized reference light beam, where full information on the spatially varying polarization state is successfully recovered by quantifying the space varying geometric phase and the contrast of interference. The proposed method enables single-shot measurement of any space varying polarization state of light from the measured interference pattern with a polarized reference beam. This approach shows considerable potential for instantaneous mapping of complex space varying polarization of light in diverse applications, such as astronomy, biomedical imaging, nanophotonics, etc., where high precision and near real-time measurement of spatial polarization patterns are desirable.

# I. INTRODUCTION

The polarization measurement [1–3] has played important roles in our understanding of the complex structure of different biological samples [3], probing the dynamics of astrophysical phenomena [4, 5], uncovering three dimensional characteristics of chemical bonds [6] and for the characterization of the complex nanomaterials and so forth [7]. Traditionally polarization is measured using the four Stokes vector elements. This however involves multiple intensity measurements which is not amenable for probing fast dynamical processes. Attempts have therefore been made to develop techniques that can instantaneously provide the full spatial polarization map, by simultaneously recording the different Stokes vector elements in multiple optical paths using combinations of polarizing beam splitters, retarders etc. [8]. However, for applications involving polarization measurements with high spatial resolution and specifically for applications where the signal to noise ratio of polarization is weak (e.g., in solar coronal magnetometry [9]), it is preferable to determine the space varying polarization in a single optical path. This follows because division of light in multiple optical paths leads to further degradation of signal to noise ratio. An alternative method is therefore highly saught after for such applications.

The concept of geometrical phase and the associated spin-orbit interaction (SOI) of light may provide a novel route for this purpose. Note that the angular momentum carried by light can be divided into spin and orbital components, which are related to circular (elliptical) polarization [10] and phase vortex [11, 12] respectively. The spin and the orbital degrees of freedom of light can get coupled under certain circumstances leading to interesting consequences like SOI [13, 14]. The SOI of light is attracting much attention due to its fundamental nature and potential applications in the development of novel spin photonic devices [7, 15]. The SOI effect has an inherent geometrical origin that relates to the evolution of geometric phase of light [15–19]. The topology of the evolution of the electromagnetic wave introduces two types of geometric phases [7, 14, 19–26]: (a) Spin redirectionBerry phase, which arises when polarized light is passed through a twisted dielectric medium [14, 23, 25]; and (b) Pancharatnam-Berry phase, which arises due to continuous change in the polarization state of the wave when polarized light propagates through an anisotropic medium [7, 19–24, 26]. An interesting manifestation of SOI is the so-called spin Hall effect of light that leads to the spin or circular polarization dependent splitting of light beam and has been observed in various optical interactions [7, 14, 16, 24]. The spin Hall effect is known to originate from the transverse (with respect to the direction of propagation) spatial or momentum gradient of either the Pancharatnam-Berry geometric phase or the spin redirection Berry phase [7, 14, 16]. For incident homogeneously polarized light beam, the spin Hall effect produced in a transversely inhomogeneous anisotropic medium that arises due to the Pancharatnam-Berry geometric phase gradient is usually larger in magnitude as compared to those produced due to the spin redirection Berry phase in isotropic medium, e.g., in tight focusing of fundamental or higher order Gaussian beams, scattering from micro, nanoscale systems, reflection, refraction at dielectric interfaces, etc. [7, 15, 24, 27].

It is known that light beam carrying inhomogeneous spatially varying polarization produces similar SOI effects due to the generation of space varying Pancharatnam-Berry geometric phase while propagating through a homogeneous anisotropic medium [24-28]. Thus, if any such spatially varying polarized light is passed through a homogeneous anisotropic medium, one can obtain information on the unknown space varying polarization state through quantification of the space varying geometric phase. In this paper, we demonstrate this useful concept by developing a simple yet elegant interferometric technique for the quantification of space varying polarization state of light. Specifically, we concentrate on single shot determination of space varying general elliptical polarization states because simultaneous retrieval of space varying ellipticity and orientation of polarization ellipse is challenging and has numerous potential applications. For this purpose, we first demonstrate the manifestation of the space varying geometric phase as spin-dependent splitting of spatially varying polarized light beam while propagating through a homogeneous anisotropic medium. This lays the foundation for our next step, in which we quantify the spatially varying polarization of light in a Mach-Zehnder interferometric arrangement through quantification of the geometric phase and the interference contrast. The demonstrated principle of single-shot determination of spatially varying polarization states of light may find useful applications for mapping complex space varying polarization patterns of remote light sources in diverse fields of physics ranging from nanophotonics [7], materials characterization [6, 29], biophotonics [3] to astronomy [4, 5].

## II. THEORY

When an *inhomogenously* polarized or spatially varying polarized Gaussian beam propagates through a *homogeneous* anisotropic medium, it acquires a space varying geometric phase. It can be shown that when such spatially polarized Gaussian beam carrying space varying geometric phase is projected to right (RCP) or left circular polarization (LCP), it exhibits opposite momentum domain shift of the centroid of the beam proportional to the spatial gradient of geometric phase (see Supporting information S1) [30]. Such spin (circular polarization)-dependent shift is similar to momentum domain spin Hall effect of light [28, 31, 32]. In order to realize this phenomenon, we consider the case of a spatially varying polarized beam generated using a twisted nematic liquid crystal-based spatial light modulator (SLM). Here, the symmetry of the polarization is broken along one of the transverse spatial co-ordinates ($\xi \rightarrow x/y$) of the beam. The evolution of polarization in SLM can be modelled using the effective Jones matrix ($J_{eff}$) as a sequential product of matrices of an equivalent linear retarder ($J_{reta}$, with effective linear retardance $\delta_{eff}$ and its orientation angle $\theta_{eff}$) and an effective optical rotator ($R$) with optical rotation $\psi_{eff}$[31, 32].

$$J_{eff} = R(\psi_{eff}) J_{reta}(\delta_{eff}, \theta_{eff})$$

$$= \begin{pmatrix} \cos\psi_{eff} & \sin\psi_{eff} \\ -\sin\psi_{eff} & \cos\psi_{eff} \end{pmatrix} \begin{pmatrix} \cos\theta_{eff} & -\sin\theta_{eff} \\ \sin\theta_{eff} & \cos\theta_{eff} \end{pmatrix} \begin{pmatrix} e^{-i\frac{\delta_{eff}}{2}} & 0 \\ 0 & e^{i\frac{\delta_{eff}}{2}} \end{pmatrix} \begin{pmatrix} \cos\theta_{eff} & \sin\theta_{eff} \\ -\sin\theta_{eff} & \cos\theta_{eff} \end{pmatrix}$$

with $\psi_{eff} = -\psi + 2\theta_{eff}$ (1)

Here, $\psi = \pi/2$ is the twist angle of the SLM. As evident from eq.1, the state of polarization of light, i.e., the ellipticity and the orientation angle of the polarization ellipse emerging from the SLM are determined by the polarization birefringence parameters, $\delta_{eff}$ and $\psi_{eff}$. These parameters in the individual SLM pixels can be modulated by changing the grey level values ($n$). Therefore, by modulating the pixels of the SLM using user-controlled grey level distributions, one can produce any desirable spatially varying polarization states of light using linearly polarized light as input. When such spatially varying polarized light beam propagates through a *homogeneous* anisotropic medium (e.g., a *half waveplate*, whose fast axis is oriented horizontally) and subsequently projected to opposite circular polarization states, a space varying geometric phase is generated.

Here, the three relevant polarization states are:

$$|A_{SVP}\rangle = \left(\cos\frac{\delta_{eff}(\xi)}{2}\cos\psi_{eff}(\xi), \quad -i\sin\frac{\delta_{eff}(\xi)}{2} - \cos\frac{\delta_{eff}(\xi)}{2}\sin\psi_{eff}(\xi)\right)^T$$

$$|B_{SVP}\rangle = \left(-i\sin\frac{\delta_{eff}(\xi)}{2} - \cos\frac{\delta_{eff}(\xi)}{2}\sin\psi_{eff}(\xi), \quad \cos\frac{\delta_{eff}(\xi)}{2}\cos\psi_{eff}(\xi)\right)^T$$

$$|C\rangle = \frac{1}{\sqrt{2}}(1, \quad \pm i)^T$$

where, $|A_{SVP}\rangle$ and $|B_{SVP}\rangle$ are the inhomogeneous spatially varying polarization state coming out of the SLM and the spatially varying polarization state after passing through the *half waveplate* respectively. The inhomogeneously polarized light coming out of the half waveplate is projected to right circular polarization (RCP,σ⁺) or left circular polarization (LCP,σ⁻) polarization states, which is denoted as $|C\rangle$. The corresponding expression for the geometric phase that is acquired due to the evolution of polarization states can be obtained using the Pancharatnam's connection as [32, 33],

$$\phi_{PB}(\xi) = \arg(\langle A_{SVP}|C\rangle\langle C|B_{SVP}\rangle\langle B_{SVP}|A_{SVP}\rangle) \qquad (2)$$

One can produce any desirable spatial gradient of geometric phase by using appropriate gradient of the grey levels $\left(\frac{dn}{d\xi=x/y}\right)$ along a chosen linear direction ($\xi \to x/y$) of the SLM [30]. The resulting spatial gradient of the geometric phase manifests as shifts in the transverse momentum distribution $\left(\frac{k_{x/y}}{2\pi}\right)$ in the Fourier domain of the Gaussian beam along the direction of the gradient, when projected to RCP and LCP states (opposite shifts for RCP and LCP, see Supporting information S1). The corresponding momentum domain spin Hall shift ($\pm\Delta k_{x/y}$) can be quantified by detecting the shift of the centroid of the Gaussian momentum distribution [32]. Experimental demonstration of this concept is presented subsequently.

Having described the space varying Pancharatnam-Berry geometric phase acquired by spatially varying polarized light beam while propagating through a homogeneous anisotropic medium, we now turn to the quantification of the spatially varying polarization state through quantification of the geometric phase information. In general, the geometric phase depends upon the orientation of the polarization ellipse with respect to the anisotropy axis of the homogeneous anisotropic medium [15, 17, 26]. Moreover, if the transmitted spatially varying polarized light beam is made to interfere with a linearly polarized reference beam, the information on the polarization ellipticity and the orientation of the polarization ellipse will also be encoded in the spatially varying contrast of the interference. Therefore, in principle, the geometric phase and the contrast information can be combined to obtain complete information on the space varying elliptical polarization (ellipticity and orientation of ellipse) using interferometric measurement with appropriate calibration of the dynamical phase of the interferometer. In order to experimentally demonstrate this concept, we take the example of the symmetry broken spatially varying polarized light beam generated by the SLM, described above. Note that unlike the previous case, the state $|C\rangle$ is not required here for retrieval of space varying polarization using interferometric determination of space varying geometric phase of light. For this specific interferometric arrangement, the space varying geometric phase is acquired in one arm of the interferometer due to the propagation of the spatially varying polarized state generated by the SLM $\left(|A_{SVP}(\xi)\rangle\right)$ through the homogeneous anisotropic medium or the *half waveplate* and subsequent evolution of the state to $|B_{SVP}\rangle$. The corresponding expression for the geometric phase is then given by,

$$\phi_{PB}(\xi) = \arg\left(\langle A_{SVP}(\xi)|B_{SVP}(\xi)\rangle\right) = \tan^{-1}\left(\frac{\tan\frac{\delta_{eff}(\xi)}{2}}{\sin\psi_{eff}(\xi)}\right) \quad (3)$$

As evident, in this case, the generated space varying geometric phase depends upon the spatial distribution of the $\delta_{eff}(\xi)$ and $\psi_{eff}(\xi)$ birefringence parameters of the SLM. The space varying polarization state 'to be reconstructed' ($|A_{SVP}(\xi)\rangle$) (which is generated using the SLM) is also determined by these polarization birefringence parameters. Thus, our job here is to determine the $\delta_{eff}(\xi)$ and $\psi_{eff}(\xi)$ birefringence parameters through interferometric quantification of the geometric phase of light, which can then be subsequently used to retrieve the space varying polarization of the light beam. Note that in principle, any arbitarary space varying polarization state coming from a distant source can be determined using this approach by quantifying the ellipticity and orientation of polarization ellipse through interferometric quantification of the geometric phase information with proper calibration of the dynamical phase of the interferometer.

It can be seen from eq. 3 that the geometric phase has two composite parameters ($\delta_{eff}(\xi)$ and $\psi_{eff}(\xi)$). Hence, that alone does not allow us to quantify them seperately. As previously noted, since the reference light beam of the interferometer is linearly polarized, the ellipticity of polarization in the sample interference arm (which is also determined by the $\delta_{eff}(\xi)$ and $\psi_{eff}(\xi)$ birefringence parameters) will affect the contrast of interference. So, with an extra measurement of contrast along with the geometric phase information, one can extract both the space varying polarization parameters ($\delta_{eff}(\xi)$ and $\psi_{eff}(\xi)$) and subsequently retrive the space varying polarization. In order to accomplish this, we proceed as follows: (1) we first determine the space varying Pancharatnam-Berry geometric phase from the interference pattern in a Mach-Zehnder configuration with appropriate calibration of the dynamical phase of the interferometric system. (2) We then quantify the spatially varying contrast of the interference fringe. The birefringence parameters extracted using suitable calibration of the dependence of the contrast on these polarization parameters of the SLM. This information is subsequently used to retrieve the spatially varying polarization state of light generated by the SLM.

### III. EXPERIMENTAL METHODS

A schematic of the experimental arrangement for observing the spin-dependent splitting of the spatially varying polarized light beam is shown in Figure1(a). Fundamental Gaussian mode of 632.8 nm line of a He-Ne laser is used to seed the system. The spatially varying polarized beam is generated using the polarization state generator (PSG) unit, which comprises of a fixed linear polarizer (P1) and a transmissive SLM. Grey level gradient ($\frac{dn}{d\xi=x/y}$= 0.0653bit μm$^{-1}$) was created in the SLM pixels along one chosen linear direction ($\xi \rightarrow$ x/y) using a range of grey level values between $n$ = 40 to 90. This choice was driven by the observed linear variations of the $\delta_{eff}(n)$ and $\psi_{eff}(n)$ parameters over this range of n values [30]. The beam is then passed through a half waveplate, which acts as the homogeneous anisotropic medium. The transmitted light beam is then sequentially analyzed for RCP and LCP basis states by projecting it to opposite circular analyzers. The circular analyzer (CA) comprises of a rotatable quarter waveplate and a linear polarizer. The beam transmitted through the circular analyzer was imaged into a CCD camera. The CCD camera was placed at the back focal Fourier plane of the Fourier transforming lens with focal length $f$. In this configuration, the recorded intensity distribution at the CCD $(x', y')$ plane (Fourier plane) represented the transverse momentum (spatial frequency) distribution $\left[\frac{k_x}{2\pi} = \frac{x'}{\lambda f}; \frac{k_y}{2\pi} = \frac{y'}{\lambda f}\right]$. The momentum domain spin Hall shift was quantified by determining the shift in the centroid of the transverse momentum distribution $\langle\frac{k_{x/y}}{2\pi}\rangle$ between the RCP and the LCP projections.

The Mach-Zehnderinterferometric arrangement for the quantification of space varying polarization is shown in Figure 1(b). The spatially varying polarized light beam generated by the SLM is used in one arm of the interferometer and a horizontally polarized Gaussian beam is used as the reference beam in the other arm. The interference pattern is captured using a CCD camera.

### IV. RESULTS AND DISCUSSIONS

*Spin-dependent splitting of SVP light beam in homogeneous anisotropic medium*

Making use of the experimental arrangement (Fig. 1(a)), spin-dependent splitting of the transverse momentum distribution $\left(\frac{k_{x/y}}{2\pi}\mu m^{-1}\right)$ of the spatially varying polarized beam is demonstrated in Figure 2. Note that before recording the spin dependent splitting of spatially varying polarized light (using grey level gradient in the SLM), measurements were taken by giving uniform grey level distribution in the SLM. No shift in the centroid of the Gaussian beam was recorded when the output beam was projected to RCP and LCP states, respectively by rotating the quarter waveplate at the detection end. Taking this measurement as a reference, similar measurements were subsequently performed on spatially varying polarized light by applying desirable grey level gradient in the SLM. The projected LCP mode ($\sigma-$, top panel) and the RCP mode ($\sigma+$, bottom panel) experienced opposite momentum domain shifts, observed as a shift of the beam centroid in the detection plane (Fig. 2(a)). This is manifested as a spin separation or spatially separated regions of opposite circular polarization states in the circular polarization descriptor Stokes Vector element $\frac{V}{I}\left[=\frac{(I_{RCP}-I_{LCP})}{(I_{RCP}+I_{LCP})}\right]$ (Fig. 2(b)). The spin (circular polarization)-dependent momentum domain beam shift corresponding to our experimental configuration was theoretically estimated using Eq. S1 and S2 of Supporting information (see Supporting information S1). The experimentally observed shift of the centroid of the Gaussian momentum distribution in the Fourier plane ($\frac{\Delta k_{x/y}}{2\pi} \sim \pm$ 13.68 x $10^{-4}$ µm$^{-1}$) is found to be in excellent agreement with the theoretically calculated momentum domain shift ($\pm 14.64 \times 10^{-4} \mu m^{-1}$). These results provide evidence that spatially varying polarized light beam generates space varying geometric phase while propagating through a homogeneous anisotropic medium, quantification of which opens-up an interesting avenue for the determination of any space varying polarization state of light.

*Interferometric determination of space varying polarization*

**Figure 3** summarizes the results of quantification of space varying polarization through quantification of Pancharatnam-Berry geometric phase using the Mach-Zehnder interferometric arrangement (Fig. 1(b)). A homogeneously polarized light beam, generated by applying uniform grey level distribution at the SLM, was first used in the sample arm to calibrate the dynamical phase of the interferometric arrangement. Interference patterns with linearly polarized reference beam were sequentially recorded using both the homogeneously polarized and the spatially varying polarized light beam passing through the *half waveplate*. The spatial variation of the phase at the CCD plane was subsequently quantified using the conventional Fourier transform method along with the phase unwrapping procedure [32–34]. The extracted unwrapped dynamical phase of the interferometer corresponding to the homogeneously polarized light beam (Fig. 3(a)), the unwrapped total phase (dynamical + geometric phase) corresponding to the spatially varying polarized light beam (Fig. 3(b)) and the geometric phase (Fig. 3(c)) that is exclusively related to the spatially varying polarization state of light, are displayed in Fig. 3(a)-3(c). As anticipated, the extracted geometric phase exhibits a spatial gradient along the direction (*y*) of the grey level gradient in the SLM. Figure 3(d) shows the calibration of interference contrast with varying optical rotation parameter $\psi_{eff}(n)$ of the SLM for different sets of homogeneously polarized beam (varying *n* = 40 to 90). Since the circular birefringence or the optical rotation parameter $\psi_{eff}$ determines the orientation of the polarization ellipse in the sample arm of the interferometer, increasing magnitude of $\psi_{eff}$ with respect to the linear polarization state that is used in the reference arm will degrade the contrast of interference. This calibration curve is subsequently utilized to quantify the spatial (*y*) variation of the $\psi_{eff}$ parameter (Fig. 3(e)) from the spatially varying contrast of interference (top panel of Fig. 3(e) and inset). Once the $\psi_{eff}(y)$ parameter is determined, the extracted spatial variation of the geometric phase (Fig. 3(c)) yields the spatial variation of the other polarization birefringence parameter, the linear retardance $\delta_{eff}(y)$ of the SLM using eq. 3 (shown in Fig. 3(f)).

As previously noted, the controlled input $\delta_{eff}$ and $\psi_{eff}$, the birefringence parameters of the SLM contain information on the ellipticity and the orientation angle of polarization ellipse, respectively of the polarization state generated by the SLM. Therefore, the retrieved spatial variation of the $\delta_{eff}(y)$ and $\psi_{eff}(y)$ parameters are used to recover the spatially varying polarization state of the light beam generated by the SLM. The corresponding results are summarized in Figure 4. The retrieved spatial variation of the polarization birefringence parameters $\delta_{eff}$ and $\psi_{eff}$ and the recovered spatially varying polarization states show excellent agreement with the controlled input parameters (Fig. 4(a)) and the corresponding spatially varying polarization state generated by the SLM (Fig. 4(b)), respectively. In order to further comprehend these experimental results, we have presented simulation of all the above steps of the experiment involving interferometric determination of geometric phase and dynamical phase of light for subsequent retrieval of space varying polarization, in Supporting information (section: S2 and Figure: S1). The presented experimental results and the corresponding results of the simulations clearly demonstrate the ability of the proposed technique for the quantification of the space varying polarizationstate of light through quantification of the space varying Pancharatnam-Berry geometric phase using a single shot interferometric measurement. The principle has been demonstrated for the general case of space varying elliptical polarization and thus it is equally applicable for space varying linear polarization states as well. As an example, simulation results of this approach for retrieving radially polarized vector beam is shown in Supporting information, section S3 (Figure S2). However, retrieval of space varying linear polarization is less challenging in general and can be achieved using other simpler approaches also.

We emphasize that in this particular scenario of space varying polarization state generated by an SLM, the demonstrated interferometric determination of the $\delta_{eff}(\xi \rightarrow x/y)$ and $\psi_{eff}(\xi \rightarrow x/y)$ parameters are equivalent to determination of the space varying ellipticity and orientation of polarization ellipse for the general case of any spatially varying polarized light from remote sources. Moreover, since in this specific case, the space varying polarization state was generated by introducing the SLM in one arm of the interferometer, it necessitated cumbersome calibration of the dependence of the contrast on the birefringence parameters of the SLM (shown in Fig. 3(d)). This may not be needed for the general case, where the extracted spatial variation of the geometric phase and the interference contrast contain sufficient information to retrieve complete information on the spatial variation of the polarization. Although the principle has been demonstrated for completely polarized light, it can also be extendedto incorporate partial polarization states by including additional calibration of the dependence of the contrast of interference on the degree of polarization of light.

## V. CONCLUSIONS

In summary, we have demonstrated a new experimental technique for the quantification of space varying polarization state of light using a single-shot interferometric measurement. This method is based on the determination of space varying Pancharatnam-Berry geometric phase that a spatially varying polarized light generates while propagating through a homogeneous anisotropic medium. It is shown that the information on the space varying geometric phase and the space varying contrast of the interference obtained using a polarized reference light beam can be combined to yield complete information on the spatially varying elliptically polarized light (ellipticity and orientation of ellipse). This principle is experimentally demonstrated in a Mach-Zehnder interferometric arrangement by recovering spatially varying polarized light generated by a spatial light modulator. With regard to the accuracy and sensitivity of the method, we would like to note that the sensitivity and accuracy of determination of space varying polarization is entirely determined by the sensitivity of the interferometric set-up for the quantification of phase. Since this method is based on quantification of phase through interferometry, it is expected to yield better sensitivity as compared to the traditional intensity-based polarization measurement methods [1–3]. This, however, remains to be rigorously evaluated. This single-shot interferometric polarimetry technique may find useful applications for probing the dynamics of a wide range of phenomena in diverse systems ranging from complex materials [6, 29], biological systems [3], to the astrophysical domain [4, 5], where high cadence and high precision measurement of spatial polarization patterns are desirable [9]. Finally, the proposed interferometric approach for spatially varying polarization measurements represents a fundamentally interesting approach with much potential. We are currently expanding our investigations toward its practical applications.

**Figures:**

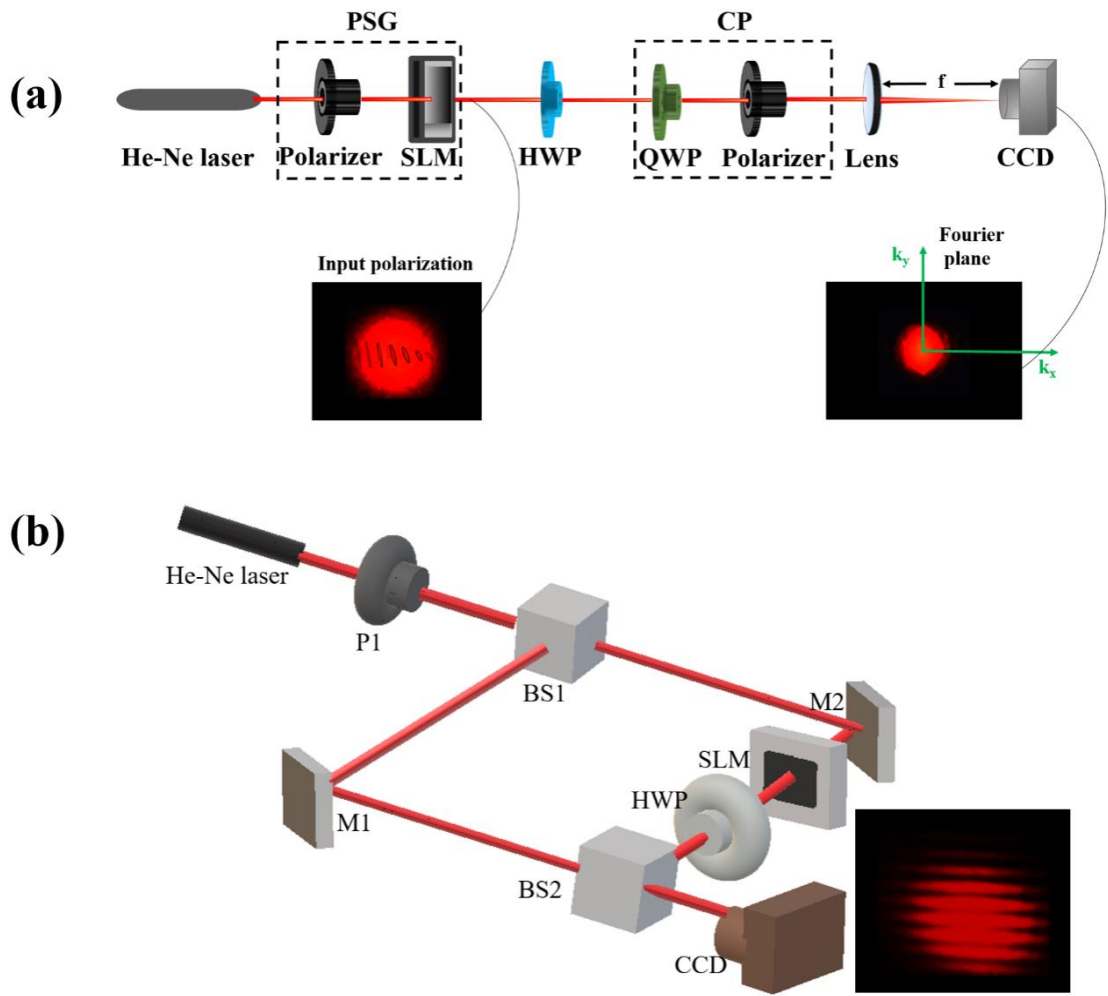

**Figure 1: (a)** A schematic of the experimental arrangement for observing spin-dependent splitting of the SVP light beam in homogeneous anisotropic medium. The PSG (polarization state generator) unit comprises of a linear polarizer followed by a spatial light modulator (SLM). A half waveplate (HWP) is used as the homogeneous anisotropic medium. For observing the spin-dependent splitting of light, the transmitted beam was analyzed via the circular analyzer comprising of a rotatable quarter waveplate(QWP) followed by a linear polarizer. **(b)** Mach-Zehnder interferometric arrangement for the quantification of space varying polarization state of light. P1:Polarizer, BS1, BS2: Beamsplitters, M1, M2: Mirrors, HWP: Half waveplate.

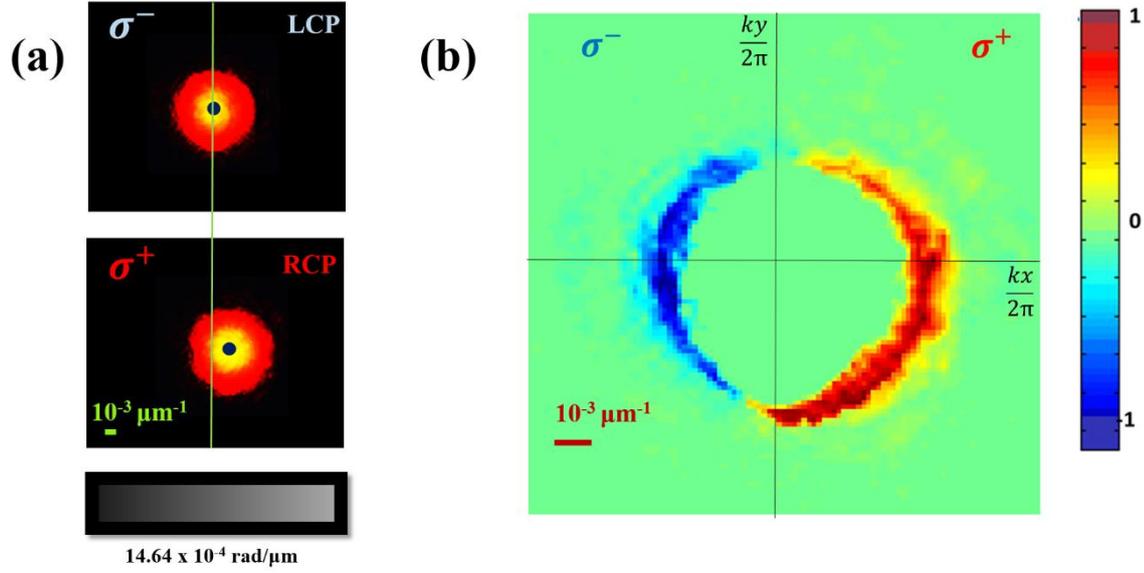

**Figure 2:** *Manifestation of space varying PB geometric phase as spin-dependent splitting of SVP light beam in homogeneous anisotropic medium.* **(a)** Transverse momentum distributions ($\frac{k_{x/y}}{2\pi}$μm$^{-1}$) of the transmitted beam for the symmetry broken SVP beam. Opposite shifts in the transverse momentum distribution of the LCP ($\sigma^-$, top panel) and RCP ($\sigma^+$, middle panel)-analyzed symmetry broken SVP beam are observed. Grey level gradient was applied along the *x*-direction in the SLM and the applied grey level values (*n*= 40-90) are displayed (bottom panel). **(b)** The corresponding distribution of the Stokes Vector element $\frac{V}{I}$.

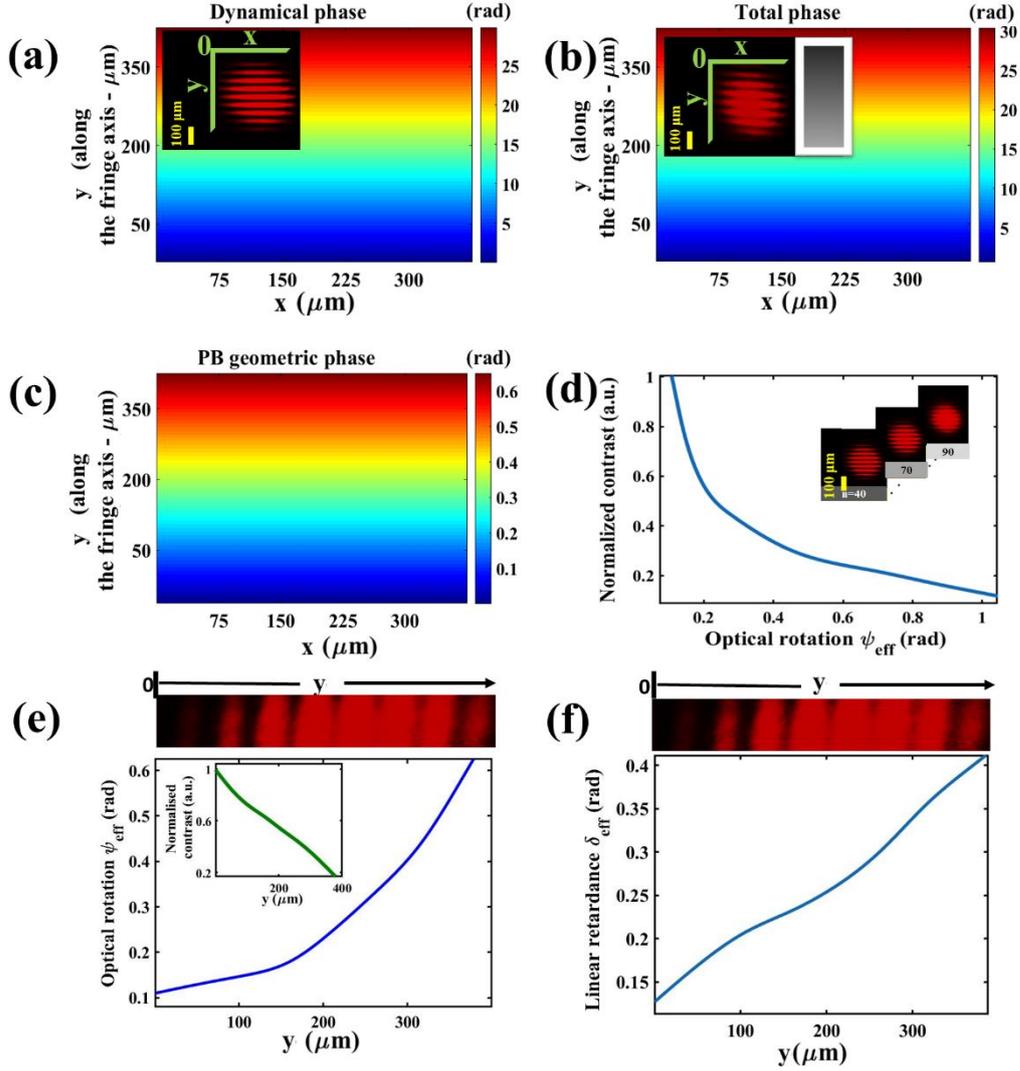

**Figure 3:** *Interferometric determination of space varying polarization through quantification of PB geometric phase.* **(a)** The dynamical phase of the interferometer extracted from the interference fringes (shown in the inset) corresponding to a homogeneously polarized light beam. The total phase **(b)** and the PB geometric phase **(c)** extracted from the interference fringe (shown in the inset) corresponding to the SVP light beam. **(d)** The dependence of the contrast of interference on varying optical rotation parameter $\psi_{eff}$ of the SLM obtained by changing the grey level values ($n=$ 40 to 90) for different sets of homogeneously polarized beam. The inset shows the corresponding interference patterns for different $n$. **(e)** The spatial (y) variation of the $\psi_{eff}$ parameter derived from the spatially varying contrast (corresponding to the fringe shown in the top panel) when the SVP light beam is used. The inset shows the corresponding spatial (y)-dependence of the contrast. **(f)** The derived spatial variation of the linear retardance parameter $\delta_{eff}(y)$ of the SLM for the SVP beam.

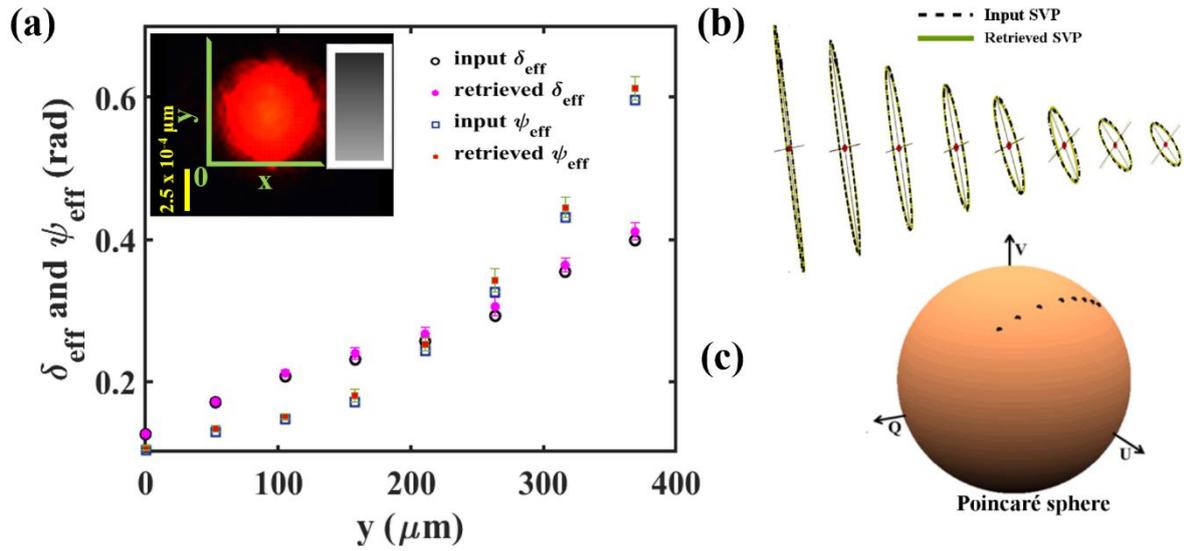

**Figure 4:** *Comparison of the retrieved space varying polarization state using the interferometric measurement with the controlled input.* **(a)** The spatial ($y$) variation of the derived (solid symbols) polarization birefringence parameters $\delta_{eff}(y)$ (circle) and $\psi_{eff}(y)$ (square) and the corresponding controlled inputs (open symbols) in the SLM. **(b)** The corresponding retrieved spatially varying polarization states (solid line) and the controlled input states (dashed line) at the SLM. **(c)** A Poincare sphere representation of the retrieved spatially varying polarization states.